\documentclass[aps,prl,twocolumn,floatfix,groupedaddress,superscriptaddress,nofootinbib,notitlepage]{revtex4-1}
\usepackage{times,bm,bbm,bbold,amssymb,amsmath,amsfonts,dsfont,cancel,graphics,graphicx,color}
\usepackage[normalem]{ulem}
\usepackage[usenames,dvipsnames]{xcolor}
\usepackage[colorlinks=true,citecolor=blue,linkcolor=blue,urlcolor=blue]{hyperref}

\DeclareFontFamily{OT1}{pzc}{}
\DeclareFontShape{OT1}{pzc}{m}{it}
              {<-> s * [1.25] pzcmi7t}{}
\DeclareMathAlphabet{\mathpzc}{OT1}{pzc}
                                 {m}{it}

\usepackage[T1]{fontenc} 
\usepackage{newtxtext,newtxmath} 

\begin{document}
\title{Mutual Information is not a Reliable Measure for Variations in Total Correlations}

\author{ S. Alipour}
\affiliation{QTF Center of Excellence, Department of Applied Physics, Aalto University, P. O. Box 11000, FI-00076 Aalto, Espoo, Finland}

\author{ S. Tuohino}
\affiliation{Faculty of Science, University of Oulu, 90570 Oulu, Finland}

\author{ A. T. Rezakhani}
\affiliation{Department of Physics, Sharif University of Technology, Tehran 14588, Iran}

\author{T. Ala-Nissila}
\affiliation{QTF Center of Excellence, Department of Applied Physics, Aalto University, P. O. Box 11000, FI-00076 Aalto, Espoo, Finland}
\affiliation{Interdisciplinary Centre for Mathematical Modelling and Department of Mathematical Sciences, Loughborough University, Loughborough, Leicestershire LE11 3TU, UK}

\begin{abstract}
Correlations disguised in various forms underlie a host of important phenomena in classical and quantum systems, such as information and energy exchanges. The quantum mutual information and the norm of the correlation matrix are both considered as proper measures of total correlations. We demonstrate that, when applied to the same system, these two measures can actually show significantly different behavior except at least in two limiting cases: when there are no correlations and when there is maximal quantum entanglement. We further quantify the discrepancy by providing analytic formulas for time derivatives of the measures for an interacting bipartite system evolving unitarily. We argue that to properly account for correlations, one should consider the full information provided by the correlation matrix (and reduced states of the subsystems). Scalar quantities such as the norm of the correlation matrix or the quantum mutual information can only capture a part of the complex features of correlations. As a concrete example, we show that in describing heat exchange associated with correlations, neither of these quantities can fully capture the underlying physics. 
\end{abstract}
\date{\today}
\maketitle

\textit{Introduction.---}With recent developments in the rapidly growing field of quantum information science, numerous novel concepts have been introduced and used for a plethora of systems and applications in physics, communications, and computation \cite{book:Nielsen-Chuang,book:wilde}, where \textit{correlations} (quantum in particular) play a key role. For example, in quantum thermodynamics \cite{goold-huber}, it is known that correlations can be a resource in energy transfer \cite{Lloyd-NoDiscordNoEner,sampaio2019}, heat and work conversion \cite{Huber-work-corr,Lutz-corr-heat,Andolina-qbattery,Sapienza-corr-ThermoResource,SciRep:ABBAMR-corr}, entropy exchange \cite{Rodrigues-Rosario-entropy-rate,Hutter-Wehner-entropy-rate,Ali-Faraj-entropy-rate}, and in the performance of quantum heat engines \cite{q-heat-eng}. Correlations can also have implications on the size of the gap in the energy spectrum of correlated systems \cite{Nielsen-ent-gap}.  It has been proven that almost all quantum states have nonclassical correlations and thus can be a useful resource \cite{Acin-almostall}; however creating such resources would in turn require thermodynamic cost \cite{Bruschi-ThermoCost-corr,Huber-ThermoCost-Corr,Friis-ThermoCost-corr,Faraj-ThermoCost-corr}. 

In almost all the existing studies, the quantum mutual information measure (QMI) has become a standard tool to quantify correlations. 
There have been extensive studies to formulate a set of natural requirements for measures of (quantum) correlations \cite{Ollivier-Zurek-discord,Vedral-discord,Modi-qcorr-criteria}, especially entanglement \cite{Plenio-Virmani,Horodeckix3}, through which it has been argued that QMI is a measure of \textit{total} correlations---whether classical or quantum---that a correlated state of a composite quantum system can have \cite{Vedral-RevModPhys2002}.  QMI has various appealing properties and also admits diverse operational interpretations, e.g., as the amount of noise required to completely destroy \textit{all} correlations \cite{Groisman-mutual-inf-noise,Anshu-mutual-inf-noise,Berta-mutual-inf-noise}. 
Notwithstanding, there also exist observations showing incompatibilities between some measures of the total, classical, and quantum correlations \cite{Hayden-EntofFormExceedMutInf,Li-EntOfForm-exceed-MutualInf,walczak-totcorr.vs.mutinf,Horodecki-relative-knowledge} as opposed to what had been largely accepted, which indicate that QMI or other measures of entanglement and quantumness of correlations may not be compatible for  \textit{ordering} quantum states based on the amount of correlations they contain \cite{Dong-chi-mutualInf}. 

In this Letter we argue that, while a nonzero QMI signals existence of correlations, it is not always a reliable and faithful measure for variations in the total correlations. This shortcoming has important consequences, e.g., for quantum thermodynamics. To examine QMI, we use the correlation matrix as a benchmark. This matrix, defined by the difference between a given quantum state and the tensor product of its reductions, naturally contains all correlations-related properties of a correlated quantum state. This matrix has already been used to define a ``correlation picture'' \cite{ULL} which enables one to obtain a dynamical equation for open quantum systems \cite{Buzek-corr-mat,book:Rivas-Huelga}, and also to study heat transfer between interacting correlated quantum systems \cite{SciRep:ABBAMR-corr}. The behavior of the correlation matrix in terms of system parameters can be considered as a reasonable reference for measuring total correlations. First we show that variations in QMI in terms of a local parameter of the system can behave differently from the variations in the correlation matrix. In particular, the correlation matrix can remain constant whereas QMI does not. This difference in behavior can be attributed to the issue of ordering quantum states \cite{ordering-Virmani, ordering-Ziman} with respect to the amount of total correlations. In addition, this feature implies that QMI may not always be a good measure to characterize total correlations. We also illustrate that an analogous discrepancy can be caused by quantum dynamics. To quantify this, we consider two coupled systems and derive exact analytic formulas for the time derivatives of QMI and the norm of the correlation matrix. Moreover, we show that neither of these measures can fully capture heat exchange between two correlated quantum systems; instead, the whole correlation matrix is required. 

\textit{QMI.---}We first consider a composite bipartite quantum system $\mathsf{SB}$ whose state is described by the density matrix $\varrho_{\mathsf{SB}}$. QMI between subsystems $\mathsf{S}$ and $\mathsf{B}$ is defined as 
\begin{align}
\mathpzc{I} = \mathbbmss{S}(\varrho_{\mathsf{SB}}\Vert\varrho_{\mathsf{S}}\otimes \varrho_{\mathsf{B}}) 
= \mathbbmss{S}(\varrho_{\mathsf{S}}\otimes \varrho_{\mathsf{B}})-\mathbbmss{S}(\varrho_{\mathsf{SB}}), \label{form-2}
\end{align}
where $\varrho_{\mathsf{S}}=\mathrm{Tr}_{\mathsf{B}}[\varrho_{\mathsf{SB}}]$ and $\varrho_{\mathsf{B}}=\mathrm{Tr}_{\mathsf{S}}[\varrho_{\mathsf{SB}}]$ are the reduced density matrices of the subsystems, $\mathbbmss{S}(\varrho)=-\mathrm{Tr}[\varrho \log \varrho]$ is the von Neumann entropy of $\varrho$, and $\mathbbmss{S}(\varrho\Vert \varrho')=\mathrm{Tr}[\varrho \log \varrho]-\mathrm{Tr}[\varrho \log \varrho']$ is the relative entropy of the two states $\varrho$ and $\varrho'$.

\textit{Correlation matrix.---}For a given density matrix $\varrho_{\mathsf{SB}}$, the correlation matrix is defined as the difference of the total state from its uncorrelated counterpart $\varrho_{\mathsf{S}}\otimes \varrho_{\mathsf{B}}$ \cite{ULL}, 
\begin{align}
\chi=\varrho_{\mathsf{SB}}-\varrho_{\mathsf{S}}\otimes \varrho_{\mathsf{B}}.
\label{chi-matrix}
\end{align}
If we assume a set of orthonormal Hermitian operator bases for the subsystems, $\{\sigma_i\}_{i=0}^{d_{\mathsf{S}}^2-1}$ and $\{\eta_j\}_{j=0}^{d_{\mathsf{B}}^2-1}$, such that $\mathrm{Tr}[\sigma_i \sigma_{i'}]=\delta_{ii'}$ and $\mathrm{Tr}[\eta_j\eta_{j'}]=\delta_{jj'}$, we obtain \cite{Dong-chi-mutualInf}
\begin{align}
\chi=\textstyle{\sum_{ij}}\left(\langle \sigma_i\otimes \eta_j\rangle_{\mathsf{SB}}-  \langle \sigma_i\rangle_{\mathsf{S}} \langle\eta_j\rangle_{\mathsf{B}}\right)\sigma_i\otimes \eta_j.
\end{align}
It can seen that the coefficients of the correlation matrix in the chosen basis are standard covariance or two-point correlation functions. Such functions are well-known measures of correlations in classical joint probability distributions. All correlation-related properties of the state $\varrho_{\mathsf{SB}}$ are inscribed in the operator $\chi$. Although for a given state $\varrho_{\mathsf{SB}}$ there is a unique $\chi$, the reverse does not necessarily hold. Since $\chi$ is a matrix, there is no reason {\it a priori} why a scalar quantity such as its
 $2$-norm $\Vert \chi\Vert_{2} \equiv \sqrt{\mathrm{Tr}[\chi^{\dag}\chi]}$ would be able to capture all the relevant features associated with total correlations in the system. In fact, we will explicitly demonstrate below that the scalar measures $\mathpzc{I}$ and $\Vert \chi\Vert_{2}$ 
can show mutually incompatible information about correlations.  



\begin{figure}[tp]
\includegraphics[width=0.7\linewidth]{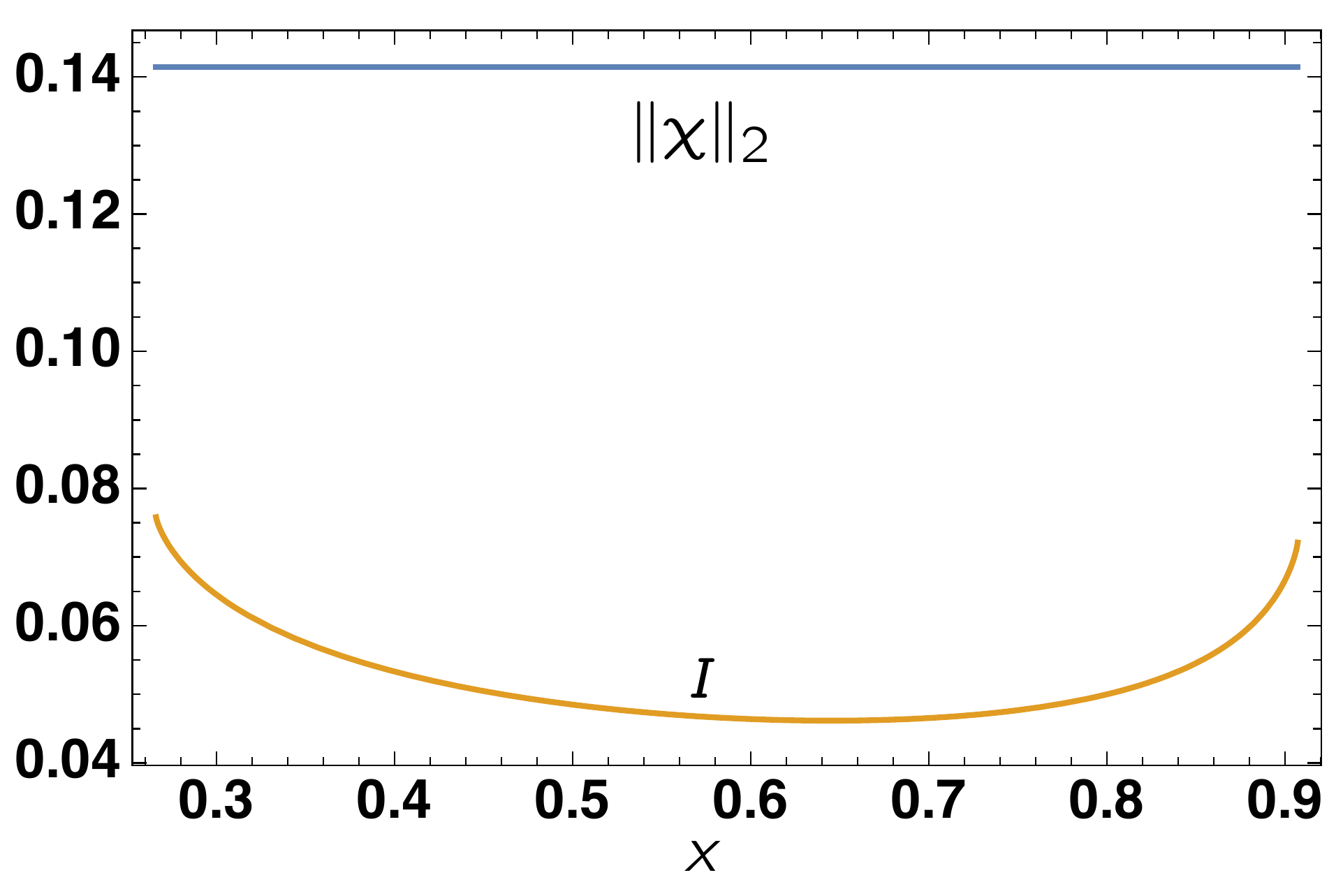}
\caption{QMI and $2$-norm of the correlation matrix for an $x$-dependent state $\varrho_{\mathsf{SB}}(x)=\varrho_{\mathsf{S}}(x)\otimes \varrho_{\mathsf{B}}(x)+\chi$ given in Example I, where the correlation matrix is parameter-independent and $\varrho_{\mathsf{S}}(x)$ and $\varrho_{\mathsf{B}}(x)$ are parameter dependent. Although the correlation matrix remains constant it can seen that QMI varies with the change of parameter and has a minimum at $x=0.64$. See text for details.}
\label{corr-vs-mutinfo}
\end{figure}

\textit{Example I. Discrepancy in the behavior between QMI and the norm of the correlation matrix.---}Let us consider a system of two interacting subsystems $\mathsf{S}$ and $\mathsf{B}$, where there is a constant correlation matrix $\chi=(|1,0\rangle\langle 0,1|+|0,1\rangle\langle 1,0|)/10$, with $\{|0\rangle,|1\rangle\}$ being the basis set of a two-dimensional space. We can choose the individual density matrices as
\begin{align*}
\varrho_{\mathsf{S}}(x)&=x|0\rangle\langle 0|+(1-x)|1\rangle\langle 1|+(|1\rangle\langle 0|+|0\rangle\langle1|)/10;\\
\varrho_{\mathsf{B}}(x)&=(1-x^2)|0\rangle\langle 0|+x^2|1\rangle\langle 1|+(|1\rangle\langle 0|+|0\rangle\langle1|)/10,
\end{align*}
for $x \in [0.27,0.9]$ and they form a proper $x$-dependent density matrix for the total system as above by $\varrho_{\mathsf{SB}}(x)=\varrho_{\mathsf{S}}(x)\otimes \varrho_{\mathsf{B}}(x)+\chi$. With these choices, while $\chi$ is constant, QMI is neither constant nor monotonic in $x$---as shown in Fig. \ref{corr-vs-mutinfo}. This result is due to the fact that, while $\Vert \chi\Vert_{2}$ measures the (scalar) magnitude of the total correlations in the system that remains constant here, QMI $\mathpzc{I}$ is a measure for \textit{relative} correlations (relative to some uncorrelated state $\varrho_{\mathsf{S}} \otimes \varrho_{\mathsf{B}}$). 

\begin{figure}[tp]
\includegraphics[width=0.85\linewidth]{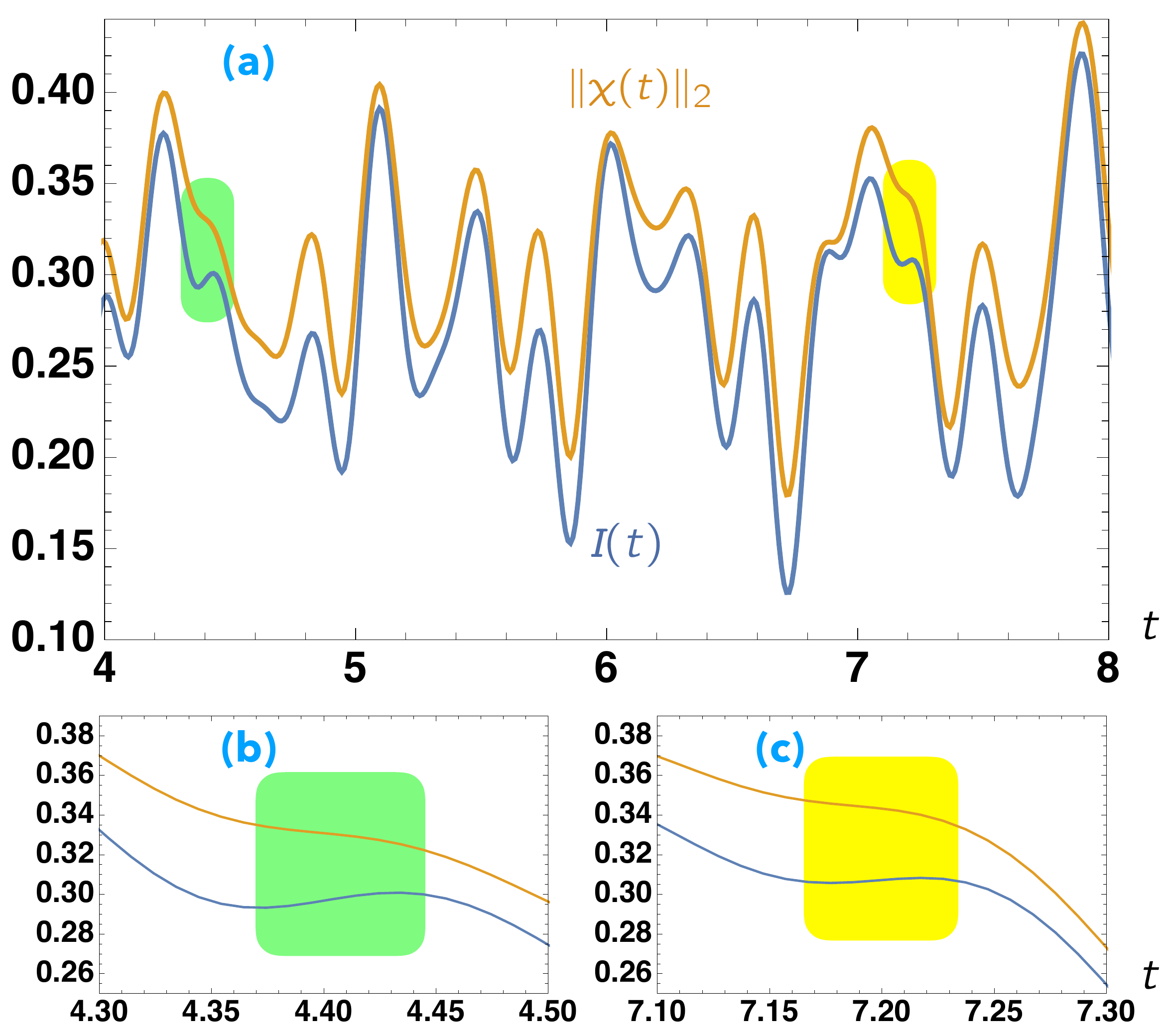}
\caption{Discrepancies in the behavior between QMI $\mathpzc{I}$ and the correlation matrix $\chi$ in a two-qubit system undergoing unitary evolution in Eaxmple II. (a) $\mathpzc{I}$ and $\Vert\chi\Vert_2$ vs time $t$. In the highlighted regions $\mathpzc{I}$ increases whereas $\Vert\chi\Vert_2$ decreases. (b) \& (c) Magnified regions of discrepancies corresponding to the green and yellow regions in (a). See the Supplementary Material for a full discussion on the discrepancies.}
\label{I-chi-discr}
\end{figure}

\textit{Example II. Discrepancy in the time-evolution of QMI and the $2$-norm of the correlation matrix.---}We now show how discrepancies 
between the scalar correlation measures can emerge due to dynamics of a bipartite system. We consider a two-qubit system with the total Hamiltonian
\begin{equation}
H_{\mathsf{SB}} = H_{\mathsf{S}} + H_{\mathsf{B}} + H_{\mathsf{I}}, 
\label{Ham}
\end{equation}
where the Hamiltonian of the first qubit ($\mathsf{S}$), the Hamiltonian of the second qubit ($\mathsf{B}$), and their interaction are given by 
\begin{align*}
H_{\mathsf{S}} =\sigma^z_\mathsf{S},\,H_{\mathsf{B}} = - \frac{1}{2}\sigma^z_\mathsf{B},\,H_{\mathsf{I}} &=\frac{5}{2}\sigma^x_\mathsf{S} \sigma^x_\mathsf{B} + \frac{1}{2} \sigma^y_\mathsf{S} \sigma^y_\mathsf{B} + \frac{9}{2}  \sigma^z_\mathsf{S} \sigma^z_\mathsf{B},
\end{align*}
with $\sigma^{\alpha}$s ($\alpha\in\{x,y,z\}$) are the Pauli matrices and $\sigma^{\pm}=(\sigma^{x}\pm i\sigma^{y})/2$. In Fig. \ref{I-chi-discr} we have plotted $\mathpzc{I}(t)$ and $\Vert \chi(t) \Vert_2$ for the evolution of a specific initial state $\varrho_\mathsf{SB}(0)$ (for details see the Supplementary Material). It can be observed that in the highlighted regions QMI displays local maxima, whereas $\Vert \chi \Vert_2$ monotonically decreases there. A closer inspection reveals that the regions of discrepancies are not restricted to the two highlighted parts; there are more time intervals at which the signs of $\mathrm{d}\mathpzc{I}(t)$ and $\mathrm{d}\Vert \chi(t) \Vert_2$ differ, as shown in the Supplementary Material. 

In general, for closed bipartite quantum systems, we can quantify the difference in time evolution between $\mathpzc{I}$ and $\Vert\chi\Vert_2$ by computing their time derivatives as  
\begin{align}
&\partial_t \mathpzc{I}=i\,\mathrm{Tr}\big[[H_{\mathsf{SB}},\chi] \log(\varrho_\mathsf{S}\otimes \varrho_\mathsf{B}) + i\chi\, \partial_{t}\log(\varrho_\mathsf{S}\otimes \varrho_\mathsf{B})\big],\label{partialI}\\
&\partial_{t}\Vert\chi\Vert_2^2=2i\,\mathrm{Tr}\big[[H_{\mathsf{SB}},\chi] \varrho_\mathsf{S}\otimes \varrho_\mathsf{B} + i \chi \, \partial_{t}(\varrho_\mathsf{S}\otimes \varrho_\mathsf{B})\big]. \label{partialchi}
\end{align}
Although the second term in Eq. (\ref{dI}) is zero, $\mathrm{Tr}[\chi \,\partial_{t}\log(\varrho_\mathsf{S}\otimes \varrho_\mathsf{B})]=0$, we keep it here for comparison. We can immediately see that the two rates are proportional only if we can approximate $\log(\varrho_\mathsf{S}\otimes \varrho_\mathsf{B})$ by $\varrho_\mathsf{S}\otimes \varrho_\mathsf{B}-\mathbbmss{I}_{\mathsf{SB}}$. This condition holds when $\varrho_{\mathsf{SB}}$ is maximally- or highly-entangled, in which case $\partial_t \mathpzc{I}(t) \approx (d_{\mathsf{SB}}/2) \partial_{t}\Vert\chi(t)\Vert_{2}^{2}$---see the Supplementary Material. In general, however, these two rates can behave in a completely different manner.


We note that despite the discrepancies demonstrated here, depending on the context and details of the physical process in question one of these measures may prove more relevant. For example,  one can expect that when information or relative entropy exchange between the subsystems is concerned QMI should provide more pertinent information, whereas in some other situations the norm of the correlation matrix may be more useful. In the following, we demonstrate that in thermodynamic processes, in particular when studying energy exchange between two correlated and interacting subsystems, the full correlation matrix---neither its norm nor QMI---is required to capture the complete physical picture.  

\textit{Binding energy vs QMI and the norm of the correlation matrix.---}Assume a bipartite quantum system with the total Hamiltonian $H_{\mathsf{SB}}$ as given in Eq. (\ref{Ham}). The binding energy $\mathbbmss{U}_{\chi}$ associated with the concurrent existence of correlations and interaction is defined as \cite{SciRep:ABBAMR-corr}
\begin{align}
\mathbbmss{U}_{\chi}=\mathrm{Tr}[\varrho_{\mathsf{SB}}\,H_{\mathsf{SB}}]-\mathrm{Tr}[\varrho_{\mathsf{S}}\otimes \varrho_{\mathsf{B}}\,H_{\mathsf{SB}}]= \mathrm{Tr}[\chi\, H_{\mathrm{I}}],
\label{U_chi}
\end{align}
where we have used $\mathrm{Tr}[\chi H_{\mathsf{S}}]=\mathrm{Tr}[\chi H_{\mathsf{B}}]=0$ (for an alternative definition of the binding energy see Ref. \cite{binding-Ali}). The thermodynamic importance of $\mathbbmss{U}_{\chi}$ is that in a time-independent closed bipartite system the heat exchange between the subsystems is compensated by the variations of $\mathbbmss{U}_{\chi}$ as
\begin{align}
\mathrm{d}\mathbbmss{Q}_{\mathsf{S}}(t) + \mathrm{d}\mathbbmss{Q}_{\mathsf{B}} (t)= - \mathrm{d}\mathbbmss{U}_{\chi}(t), 
\label{heat}
\end{align}
where $\mathrm{d}\mathbbmss{Q}_{\mathsf{S}} = \mathrm{Tr}[\mathrm{d}\varrho_{\mathsf{S}}\,\widetilde{H}_{\mathsf{S}}]$ with the effective Hamiltonian $\widetilde{H}_{\mathsf{S}}=H_{\mathsf{S}}+\mathrm{Tr}_{\mathsf{B}}[\varrho_{\mathsf{B}} \, H_{\mathrm{I}}]$ and similarly for $\mathsf{B}$. 
  
From Eq. (\ref{U_chi}) it is evident that the change of $\mathbbmss{U}_{\chi}$ is purely because of the change of $\chi$. Now we argue that neither of the quantities $\mathpzc{I}$ or $\Vert \chi\Vert_{2}$ alone suffices to explain how $\mathbbmss{U}_{\chi}$ changes---hence they are insufficient to explain how heat is exchanged between the subsystems. To do this, we derive two equivalent expressions for $\mathrm{d}\mathbbmss{U}_{\chi}$ (for the details see the Supplementary Material). The first expression is given by
\begin{align}
\mathrm{d}\mathbbmss{U}_{\chi}(t)=&\mathrm{Tr}\big[\widehat{\chi}(t)\, H_{\mathrm{I}}\big]\,\mathrm{d} \Vert\chi(t)\Vert_2+\mathrm{Tr}\big[\mathrm{d}\widehat{\chi}(t)\, H_{\mathrm{I}}\big]\,\Vert\chi(t)\Vert_2,
\label{dUchi-dchi}
\end{align}
where $\widehat{\chi}(t)=\chi(t)/\Vert \chi(t)\Vert_2$ is the normalized correlation matrix. This implies that $\mathrm{d}\mathbbmss{U}_{\chi}$ cannot be described by $\mathrm{d}\Vert\chi(t)\Vert_2$ alone due to the second term on the right-hand side.  
For the second expression, we assume a time-independent reference (thermal) state $\varrho_{\mathsf{S}}^{\star} \otimes \varrho_{\mathsf{B}}^{\star}=e^{-\beta H_\mathsf{S}}\otimes e^{-\beta H_\mathsf{B}}/\mathrm{Tr}[e^{-\beta H_\mathsf{S}}\otimes e^{-\beta H_\mathsf{B}}]$, with $\beta$ being the inverse temperature, to obtain
\begin{align}
\mathrm{d}\mathbbmss{U}_{\chi}(t)=&-(1/\beta)\,\mathrm{d}\mathpzc{I}(t)-(1/\beta)\,\mathrm{d}\mathbbmss{S}\big(\varrho_\mathsf{S}(t)\otimes \varrho_{\mathsf{B}}(t)\Vert \varrho_\mathsf{S}^{\star}\otimes \varrho_{\mathsf{B}}^{\star}\big)\nonumber\\
&-\mathrm{Tr}[\mathrm{d}\big(\varrho_{\mathsf{S}}(t)\otimes \varrho_{\mathsf{B}}(t)\big) H_{\mathrm{I}}].
\label{dI}
\end{align}
This relation also shows that due to the extra terms on the right-hand side, $\mathrm{d}\mathbbmss{U}_{\chi}$ is not necessarily monotonic in $\mathrm{d}\mathpzc{I}$. 

We can obtain an interesting additional result by assuming $\varrho_{\mathsf{SB}}(0)=\varrho_{\mathsf{S}}^{\star} \otimes \varrho_{\mathsf{B}}^{\star}$
and integrating Eq. \eqref{dI}, which gives
\begin{align}
I(t)=&-\beta~\mathrm{Tr}[\chi(t)~H_{\mathrm{I}}]-\mathbbmss{S}\big(\varrho_\mathsf{S}(t)\otimes \varrho_{\mathsf{B}}(t)\Vert \varrho_\mathsf{S}(0)\otimes \varrho_{\mathsf{B}}(0)\big)\nonumber\\
&-\beta~\mathrm{Tr}[\Delta_t(\varrho_{\mathsf{S}}\otimes \varrho_{\mathsf{B}}) H_{\mathrm{I}}],
\end{align}
where $\Delta_t(\varrho_{\mathsf{S}}\otimes \varrho_{\mathsf{B}})=\varrho_{\mathsf{S}}(t)\otimes \varrho_{\mathsf{B}}(t)-\varrho_{\mathsf{S}}(0)\otimes \varrho_{\mathsf{B}}(0)$. By using $\mathbbmss{S}\big(\varrho_\mathsf{S}(t)\otimes \varrho_{\mathsf{B}}(t)\Vert \varrho_\mathsf{S}(0)\otimes \varrho_{\mathsf{B}}(0)\big)\geqslant 0$, $\mathrm{Tr}[A B]\leqslant \Vert A\Vert_1 \Vert B \Vert$ (with $\Vert X\Vert=\sup_{\Vert |v\rangle\Vert=1}|\langle v|X|v\rangle|$ and $\Vert X\Vert_{1}=\mathrm{Tr}[\sqrt{X^{\dag}X}]$), and $\Vert \Delta_t \varrho_{\mathsf{SB}}\Vert_1 \leqslant 2$, we can get the bound
\begin{align}
\mathpzc{I}(t) \leqslant &2 \beta \Vert H_{\mathrm{I}}\Vert.
\end{align}
In generic systems where interactions are short-range and local, $\Vert H_{\mathsf{I}}\Vert$ is proportional to the ``area'' of subsystem $\mathsf{S}$---the number of boundary particles of $\mathsf{S}$ that interact with $\mathsf{B}$. Our bound thus implies an ``area law'' for QMI \cite{area-law}. Note that our derivation did not require thermal equilibrium for the state or Markovianity for its subsystem dynamics. It only requires that the initial state of the composite system is an uncorrelated thermal state of the two systems.

\textit{Summary and conclusions.---}We have shown how significant discrepancies can arise in a bipartite quantum state in the behavior between two standard measures of total correlations, namely, the quantum mutual information measure and the norm of the correlation matrix. In particular, we have demonstrated by an example that while the correlation matrix is constant and only reduced density matrices are parameter-dependent, the quantum mutual information features variations with the parameter of the local states. Moreover, we have shown that discrepancies can emerge also in the case of a bipartite system undergoing unitary evolution. We have considered the case of two coupled qubits and argued that while the norm of the correlation matrix increases (decreases) in time, the quantum mutual information may decrease (increase). This feature implies that if we consider the norm of the correlation matrix as a measure for the amount of (total) correlations within a composite quantum state, the quantum mutual information may not work well for ordering quantum states in terms of their correlations. We have further quantified the differences between the two quantities by deriving analytic expressions for their time derivatives for a general case and showing where these may agree. As an additional result, we have also obtained an upper bound on quantum mutual information, which leads to an area law that the quantum mutual information scales with the area of the system. This result holds in general except for assuming an uncorrelated thermal initial state for the system and its bath. 

\textit{Acknowledgement.---}This work was supported in part by the Academy of Finland's Center of Excellence program QTF Project 312298, Aalto University's  AScI Visiting Professor Funding, and Sharif University of Technology's Office of Vice President for Research and Technology. 


%

\onecolumngrid
\pagebreak
\appendix
\begin{center}
\textbf{Supplemental Material}
\end{center}
\section{I. Details of Example II}
\label{Ex.II-detailes}

\textit{The total system initial state.---}In Example II we have assumed the following initial state for the two-qubit system in the computational basis ($|0\rangle =(1\,\,0)^{T}$ and $|1\rangle =(0\,\,1)^{T}$):
\begin{equation} \label{eq:192}
\varrho_{\mathsf{SB}}(0) =
\begin{pmatrix}
    0.403041 & -0.181049 - 0.038525 i & 0.012466 + 0.12214 i & -0.044462 + 0.058024 i \\
    -0.181049 + 0.038525 i & 0.314013 & 0.025204 - 0.101876 i & 0.053753 + 0.030605 i \\
    0.012466 - 0.12214 i & 0.025204 + 0.101876 i & 0.065777  & -0.018686 + 0.024092 i \\
    -0.044462 - 0.058024 i & 0.053753 - 0.030605 i & -0.018686 - 0.024092 i & 0.217169
\end{pmatrix}.
\end{equation}

\begin{figure}[bp]
\includegraphics[width=0.8\linewidth]{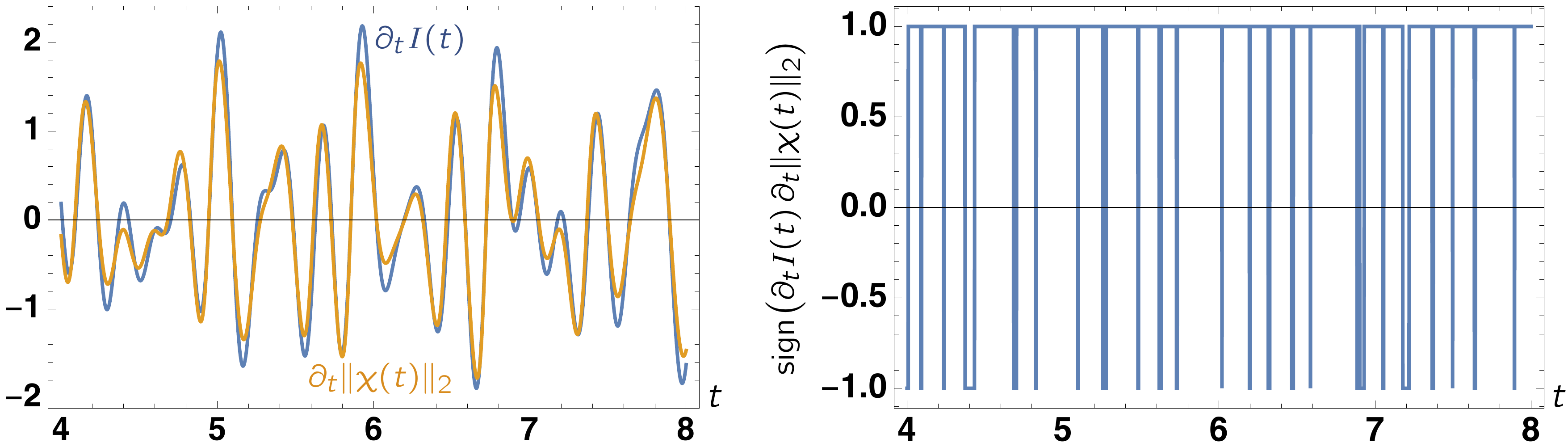}
\caption{[Left]: $\partial_{t}\mathpzc{I}$ and $\Vert\chi \Vert_{2}$ vs $t$ in example II. [Right]: Comparing the signs of $\partial_{t}\mathpzc{I}$ and $\partial_{t}\Vert \chi\Vert_{2}$. Discrepancy in the signs $\partial_{t}\mathpzc{I}$ and $\partial_{t}\Vert\chi \Vert_{2}$ is seen at the time intervals where $\mathrm{sign}\big(\partial_{t}\mathpzc{I}(t)\, \partial_{t}\Vert\chi(t) \Vert_{2})=-1$.}
\label{fig-dI-dchi-sign}
\end{figure}

\textit{Time intervals of opposite signs for $\partial_{t}\mathpzc{I}$ and $\partial_{t}\Vert\chi\Vert_2$.---}Time derivatives of $\mathpzc{I}(t)$ and $\Vert\chi(t)\Vert_{2}$ are shown in Fig. \ref{fig-dI-dchi-sign}(a). In Fig. \ref{fig-dI-dchi-sign}(b) we have depicted the sign of the derivatives and obtained the time intervals within which $\mathrm{sign}\big(\partial_t \mathpzc{I}(t)\big)\neq \mathrm{sign} \big(\partial_t \Vert\chi(t)\Vert_{2}\big)$. In the left green part in Fig. \ref{I-chi-discr}(a), the discrepancy interval is when $t \in [4.371, 4.432]$, and in the right yellow part the discrepancy is seen for $t \in [7.177, 7.218]$. The rest of the time intervals with discrepancies, given a $4\times 10^{-5}$ time-step in the numerical simulation, are as follows: $t \in \{[4.000, 4.008] \cup
[4.088, 4.093] \cup
[4.234, 4.238] \cup
[4.371, 4.432] \cup
[4.682, 4.702] \cup
[4.824, 4.829] \cup
[5.094, 5.097] \cup
[5.256, 5.275] \cup
[5.480, 5.484] \cup
[5.616, 5.628] \cup
[5.727, 5.732] \cup
[6.018, 6.019] \cup
[6.194, 6.198] \cup
[6.315, 6.325] \cup
[6.465, 6.476] \cup
[6.585, 6.587] \cup
[6.885, 6.887] \cup
[6.901, 6.931] \cup
[7.051, 7.056] \cup
[7.177, 7.218] \cup
[7.365, 7.374] \cup
[7.496, 7.497] \cup
[7.636, 7.643] \cup
[7.893, 7.895]\}$. These intervals can be seen in Fig. \ref{fig-dI-dchi-sign}(b).

\section{II. Derivation of the expressions for $\partial_t \mathpzc{I}$ and $\partial_t \Vert\chi\Vert_2^2$ [Eqs. (\ref{partialI}) and (\ref{partialchi})]}
\label{eq-1-detailes}

Since entropy is invariant under unitary evolution, $\mathbbmss{S}(\varrho_{\mathsf{SB}}(t))=\mathbbmss{S}(\varrho_{\mathsf{SB}}(0))$ is time-independent. Hence $\partial_t\mathbbmss{S}_{\mathsf{SB}}(t)=0$. This yields the rate of charge of QMI as 
\begin{align}
\partial_t \mathpzc{I} &\overset{(\ref{form-2})}{=} \partial_t \mathbbmss{S}_{\mathsf{S}}+\partial_t \mathbbmss{S}_{\mathsf{B}}\\
&=-\mathrm{Tr}\big[\partial_t(\varrho_{\mathsf{S}}\otimes \varrho_{\mathsf{B}}) \log(\varrho_{\mathsf{S}}\otimes \varrho_{\mathsf{B}})\big]\nonumber\\
&\overset{(\ref{chi-matrix})}{=}-\mathrm{Tr}\big[\partial_t(\varrho_{\mathsf{SB}}-\chi) \log(\varrho_{\mathsf{S}}\otimes \varrho_{\mathsf{B}})\big]\nonumber\\
&=i\, \mathrm{Tr}\big[[H_{\mathsf{SB}},\varrho_{\mathsf{SB}} ]\log(\varrho_{\mathsf{S}}\otimes \varrho_{\mathsf{B}})\big]+\mathrm{Tr}\big[\partial_t \chi \log(\varrho_{\mathsf{S}}\otimes \varrho_{\mathsf{B}})\big]\nonumber\\
&=i\,\mathrm{Tr}\big[[H_{\mathsf{SB}},\varrho_{\mathsf{S}}\otimes \varrho_{\mathsf{B}}+\chi ]\log(\varrho_{\mathsf{S}}\otimes \varrho_{\mathsf{b}})\big]+\mathrm{Tr}\big[\partial_t \chi \log(\varrho_{\mathsf{S}}\otimes \varrho_{\mathsf{B}})\big]\nonumber\\
&=i\, \mathrm{Tr}\big[[H_{\mathrm{I}},\chi]\log(\varrho_{\mathsf{S}}\otimes \varrho_{\mathsf{B}})\big].
\end{align}
In the last line we have used $\mathrm{Tr}\big[\partial_t \chi \log(\varrho_{\mathsf{S}}\otimes \varrho_{\mathsf{B}})\big]=0$, $\mathrm{Tr}\big[[H_{\mathsf{SB}},\varrho_{\mathsf{S}}\otimes \varrho_{\mathsf{B}}]\log(\varrho_{\mathsf{S}}\otimes\varrho_{\mathsf{B}})\big]=0$, $\mathrm{Tr}\big[[H_{\mathsf{S}},\chi]\log\varrho_{\mathsf{S}}\big]=0$, and $\mathrm{Tr}\big[[H_{a},\chi] \log\varrho_{b}\big]=0$ for $a,b \in \{\mathsf{S},\mathsf{B}\}$.

For $\partial_t \Vert\chi\Vert_2^2$ we note that 
\begin{align}
\partial_t \Vert\chi\Vert_2^2&=\partial_t \mathrm{Tr}[\chi^2]\nonumber\\
&=2\, \mathrm{Tr}[\partial_t \chi\,\chi]\nonumber\\
&=2\, \mathrm{Tr}[\partial_t (\varrho_{\mathsf{SB}}-\varrho_{\mathsf{S}}\otimes \varrho_{\mathsf{B}})\chi]\nonumber\\
&=2i\, \mathrm{Tr}\big[[H_{\mathsf{SB}},\chi]\varrho_{\mathsf{S}}\otimes \varrho_{\mathsf{B}}\big]-2\, \mathrm{Tr}[\partial_t(\varrho_{\mathsf{S}}\otimes \varrho_{\mathsf{B}})\chi].
\end{align}

\section{III. Comparing Eqs. \eqref{partialI} and \eqref{partialchi}}
\label{levy}

In Eq. (\ref{partialI}) we can replace $\varrho_{\mathsf{S}} \otimes \varrho_{\mathsf{B}}$ with $d_{\mathsf{SB}}\,\varrho_{\mathsf{S}} \otimes \varrho_{\mathsf{B}}$, without any effect on the final result. Similarly, in Eq. (\ref{partialchi}) we can replace $\varrho_{\mathsf{S}} \otimes \varrho_{\mathsf{B}}$ with $\varrho_{\mathsf{S}} \otimes \varrho_{\mathsf{B}}-\mathbbmss{I}_{\mathsf{SB}}/d_{\mathsf{SB}}$. Thus
\begin{align}
&\partial_t \mathpzc{I}=i\,\mathrm{Tr}\big[[H_{\mathsf{SB}},\chi] \log(d_{\mathsf{SB}}\,\varrho_\mathsf{S}\otimes \varrho_\mathsf{B}) + i\chi\, \partial_{t}\log(d_{\mathsf{SB}}\,\varrho_\mathsf{S}\otimes \varrho_\mathsf{B})\big],\\
&(d_{\mathsf{SB}}/2)\partial_{t}\Vert\chi\Vert_2^2=i\,\mathrm{Tr}\big[[H_{\mathsf{SB}},\chi] \big(d_{\mathsf{SB}}\,\varrho_\mathsf{S}\otimes \varrho_\mathsf{B} -\mathbbmss{I}_{\mathsf{SB}}\big) + i \chi \, \partial_{t}\big(d_{\mathsf{SB}}\,\varrho_\mathsf{S}\otimes \varrho_\mathsf{B} -\mathbbmss{I}_{\mathsf{SB}}\big)\big]. 
\end{align}
These two expressions would be equal if we can approximate $\log(d_{\mathsf{SB}}\,\varrho_\mathsf{S}\otimes \varrho_\mathsf{B})\approx d_{\mathsf{SB}}\,\varrho_\mathsf{S}\otimes \varrho_\mathsf{B} -\mathbbmss{I}_{\mathsf{SB}}$. From $\log A\approx A-\mathbbmss{I}$, for a positive operator $A$ with $\Vert A-\mathbbmss{I}\Vert \ll 1$, we can simplify this condition as
\begin{equation}
\Big\Vert\varrho_\mathsf{S}\otimes \varrho_\mathsf{B} -\frac{\mathbbmss{I}_{\mathsf{S}}}{d_{\mathsf{S}}} \otimes \frac{\mathbbmss{I}_{\mathsf{B}}}{d_{\mathsf{B}}} \Big\Vert \ll \frac{1}{d_{\mathsf{SB}}}.
\label{eq-cond}
\end{equation}
In other words, 
\begin{equation}
\varrho_\mathsf{S}(t)\otimes \varrho_\mathsf{B}(t) \approx \frac{\mathbbmss{I}_{\mathsf{S}}}{d_{\mathsf{S}}} \otimes \frac{\mathbbmss{I}_{\mathsf{B}}}{d_{\mathsf{B}}} \,\Rightarrow\, \partial_t \mathpzc{I}(t) \approx \frac{d_{\mathsf{SB}}}{2} \partial_{t}\Vert\chi(t)\Vert_{2}^{2}.
\end{equation}
Condition (\ref{eq-cond}) is satisfied when $\varrho_{\mathsf{SB}}$ is a maximally- or highly-entangled state.

\section{IV. Derivation of Eq. \eqref{dI}}
\label{eq-2-detailes}

The relation between the time derivative of QMI and heat exchange can be obtained through the following:
\begin{align}
\mathrm{d}I=&\mathrm{d} \mathbbmss{S}_{\mathsf{S}}+\mathrm{d} \mathbbmss{S}_{\mathsf{B}}\nonumber\\
=&-\mathrm{Tr}\big[\mathrm{d}\varrho_{\mathsf{S}} \log\varrho_{\mathsf{S}}\big]-\mathrm{Tr}\big[\mathrm{d}\varrho_{\mathsf{B}} \log\varrho_{\mathsf{B}}\big]\nonumber\\
=&-\big(\mathrm{Tr}\big[\mathrm{d}\varrho_{\mathsf{S}} \log\varrho_{\mathsf{S}}\big]-\mathrm{Tr}\big[\mathrm{d}\varrho_{\mathsf{S}} \log\varrho^{\star}_{\mathsf{S}}\big]\big)-\mathrm{Tr}\big[\mathrm{d}\varrho_{\mathsf{S}} \log\varrho^{\star}_{\mathsf{S}}\big]-\big(\mathrm{Tr}\big[\mathrm{d}\varrho_{\mathsf{B}} \log\varrho_{\mathsf{B}}\big]-\mathrm{Tr}\big[\mathrm{d}\varrho_{\mathsf{B}} \log\varrho^{\star}_{\mathsf{B}}\big]\big)-\mathrm{Tr}\big[\mathrm{d}\varrho_{\mathsf{B}} \log\varrho^{\star}_{\mathsf{B}}\big]\nonumber\\
=&-\mathrm{d}\mathbbmss{S}\big(\varrho_{\mathsf{S}}\Vert \varrho^{\star}_{\mathsf{S}}\big)+\beta\mathrm{Tr}\big[\mathrm{d}\varrho_{\mathsf{S}} H_{\mathsf{S}}\big]-\mathrm{d}\mathbbmss{S}\big(\varrho_{\mathsf{B}}\Vert \varrho^{\star}_{\mathsf{B}}\big)+\beta\mathrm{Tr}\big[\mathrm{d}\varrho_{\mathsf{B}} H_{\mathsf{B}}\big]\nonumber\\
=&-\mathrm{d}\mathbbmss{S}\big(\varrho_\mathsf{S}\otimes \varrho_{\mathsf{B}}\Vert \varrho_\mathsf{S}^{\star}\otimes \varrho_{\mathsf{B}}^{\star}\big)-\beta~\mathrm{Tr}[\mathrm{d}\chi(t) H_{\mathrm{I}}]-\beta~\mathrm{Tr}[\mathrm{d}\big(\varrho_{\mathsf{S}}\otimes \varrho_{\mathsf{B}}(t)\big) H_{\mathrm{I}}],
\end{align}
where $\varrho_{\mathsf{S}}^{\star}=e^{-\beta H_{\mathsf{S}}}/\mathrm{Tr}[e^{-\beta H_{\mathsf{S}}}]$, $\varrho_{\mathsf{B}}^{\star}=e^{-\beta H_{\mathsf{S}}}/\mathrm{Tr}[e^{-\beta H_{\mathsf{B}}}]$, and in the last line we have used the identity in Eq. \eqref{U_chi}.

\end{document}